\documentstyle[11pt,aaspp4]{article}

\lefthead{Smith et al.}
\righthead{Boron \& Oxygen} 

\begin{document}

\title{The Observed Trend of Boron and Oxygen in Field Stars of the Disk}

\author{Verne V. Smith}
\affil{Department of Physics, University of Texas at El Paso, El Paso, 
TX 79968  USA, and McDonald Observatory, University of Texas at Austin,
Austin, TX 78712 USA; verne@barium.physics.utep.edu}

\author{Katia Cunha}
\affil{Observat\'orio Nacional - CNPq, Rua General Jos\'e Cristino 77, 20921-400 S\~ao Crist\'ov\~ao, RJ, Brazil; katia@on.br}

\author{Jeremy R. King}
\affil{Department of Physics, University of Nevada, Las Vegas, 
4505 South Maryland Parkway, Las Vegas, NV 89154-4002 USA; 
jking@bartoli.physics.unlv.edu}

\begin{abstract}

Oxygen abundances are derived in a sample of 13 field F and G dwarfs or
subgiants with metallicities in the range of -0.75 $\le$ [Fe/H] $\le$
+0.15.  This is the same sample of stars for which boron abundances have
been derived earlier from archived spectra obtained with the Hubble Space
Telescope.  Only the weak [O I] $\lambda$6300\AA\ and  O I $\lambda$6157\AA\
and $\lambda$6158\AA\ lines have been used to determine O abundances.  It
is argued that, over the range of temperature and metallicity spanned by
the program stars, these [O I] and O I lines provide accurate oxygen
abundances, largely free from non-LTE or 1D model atmosphere effects.
The results for oxygen are combined with the boron abundances published
previously to define a boron versus oxygen abundance for field disk
stars: the relation log(B/H) + 12. = log $\epsilon$(B)=
1.39$\pm$0.08log $\epsilon$(O) - 9.62$\pm$1.38 is obtained.  The slope
of m$_{\rm BO}$=1.39 (in log-log abundance by number coordinates) indicates
that, in the disk, the abundance of B, relative to O, is intermediate
between primary and secondary production (hybrid behavior).  The slope
found here for log $\epsilon$(B) versus log $\epsilon$(O) is identical,
within the uncertainties, to that found by previous investigators for
log $\epsilon$(Be) versus log $\epsilon$(O), where m$_{\rm BeO}$=1.45.
The two relations of B and Be versus O result in essentially solar B/Be
ratios for field disk stats.  A comparison of the results here
for B--O in the disk to B--O in the halo (with B abundances taken
from the literature) reveals that, if [O/Fe] in the halo is nearly
constant, or undergoes only a gentle increase with decreasing [Fe/H],
then boron behaves as a primary element relative to oxygen.  In such a
case, there is a transition from N(B) $\alpha$ N(O) in the halo, to
N(B) $\alpha$ N(O)$^{1.4}$ in the disk.  On the other hand, if [O/Fe]
increases substantially in the halo (such that [O/Fe] $\alpha$ -0.4[Fe/H]),
as suggested by some studies of the 3100-3200\AA\ electronic OH lines, 
then there is no significant difference between the behavior of B--O
in the halo compared to the disk (i.e. N(B) $\alpha$ N(O)$^{1.4}$).

\end{abstract}

\keywords{Galaxy: abundances --- stars: abundances}
\newpage
\section{Introduction}

The light elements lithium, beryllium, and boron fall, in mass,
between the elements produced primarily as products of Big Bang
nucleosynthesis, H and He, and those produced by stellar nucleosynthesis
(elements heavier than $^{12}$C).  These three elements consist of
the five stable nuclei $^{6}$Li,
$^{7}$Li, $^{9}$Be, $^{10}$B, and $^{11}$B. The origins of these 
isotopes are believed to be understood qualitatively: the Big Bang
produced some $^{7}$Li (perhaps 10 percent of the total current
Galactic disk abundance). Spallation interactions between cosmic 
rays and ambient gas in the interstellar medium provide some amounts
of all five of these light stable nuclei. A variation of cosmic-ray
production includes fusion reactions of the form $\alpha$ plus $\alpha$
between cosmic ray $\alpha$-particles and interstellar $^{4}$He to
produce some $^{6}$Li and $^{7}$Li. Stellar nucleosynthesis also plays
a role in the synthesis of $^{7}$Li and $^{11}$B; neutrino-induced
spallation in core-collapse type II supernovae (SN II) may produce 
some $^{7}$Li and $^{11}$B, while hot bottom burning in asymptotic
giant branch (AGB) stars is known to create $^{7}$Li.

With so many processes involved in light-element synthesis, the
chemical evolution of Li, Be and B requires substantial input
from both observations and theory. Within the last 5-10 years,
the addition of new observations of boron from HST, of beryllium
from large ground-based telescopes, and the inclusion of stellar
and ISM observations of $^{6}$Li, has resulted in more emphasis 
on quantitative comparisons of yields from the various production
mechanisms of the theory. 
In this paper, the focus is on a better definition of the relationship
of boron versus oxygen for the more metal-rich regime from [Fe/H]$\sim$-0.75
to +0.15. 

Most previous studies of boron with metallicity have used iron
as the fiducial metallicity indicator.  This is due, in part, to the
ease with which Fe abundances can be measured in stars, large numbers
of Fe I and Fe II lines are easily measurable over a range of
stellar effective temperatures and metallicities.
Although easy to observe and analyze, iron is perhaps not the best choice
as a benchmark against which to measure boron: the nucleosynthetic origin 
of Fe, above [Fe/H]$\ge$-1.0, arises both from SN II and SN Ia, as
well as the fact that Fe plays essentially no role in the spallative
production of the light elements. 

Oxygen, on the other hand, is
more directly related to B production: boron produced by the $\nu$-process should scale with 
oxygen (as both the $\nu$-process and O-production result from SN II);
spallation reactions, which produce $^{10}$B and $^{11}$B, result to
a large degree from O, either as energetic protons spallating ISM 
$^{16}$O, or accelerated oxygen being spalled from ambient ISM H and He.
But oxygen is more difficult to observe than iron. The choice
of available lines is more limited and reliability of these lines
remains a topic of active research and discussion. There are basically
four ways to measure oxygen abundances in stars with T$_{eff}$ $\le$
7000K (which corresponds to the main-sequence turn-off for the lower-mass
old disk or halo stars): 1- the forbidden [O I] lines at $\lambda$ 6300\AA\
and $\lambda$ 6363\AA; 2- permitted O I lines, such as the well-known
near-IR triplet lines at $\lambda$7771-7775\AA\ (which are quite strong
at solar-like metallcities and temperatures), or weaker O I lines, such
as those found near 
$\lambda$6156-6158\AA; 3- OH from electronic transitions
in the near-UV region of 3100-3200\AA; 4- OH from vibration-rotation 
transitions in the IR near 1.6 $\mu$m and 3.4 $\mu$m. All of these
various oxygen determinants have good and bad points associated with
them, and there is an extensive body of literature on this topic; a recent
summary of ideas concerning the pros and cons of the different ways
to measure oxygen is found in the review by Lambert (2001).

Some of the major points concerning the derivation of oxygen abundances
can be summarized as follows: the zero-volt
[O I] lines should be reliable O-abundance indicators in the cool
stars, as most of the oxygen atoms will be neutral in the ground
state. However, these lines are fairly weak, requiring relatively
high-S/N spectra, and will not be observable at all for the 
more metal-poor stars. 
The permitted O I lines at 
7771-7775\AA\ are strong lines, and thus detectable even in
very metal-poor stars, however, they arise from $\chi$=9.14 eV, and
thus only represent a small fraction of total O I atoms.  It has been
argued that these lines may be affected by non-LTE, as well as
being sensitive to small scale
temperature inhomogeneities in a star's atmosphere, i.e. granulation 
(Kiselman 1991, 1993; Kiselman \& Nordlund 1995).
The size of this effect and its variation with temperature,
gravity, and metallicity remain a topic of some controversy. The other
permitted lines near $\lambda$6156-6158\AA\ are very weak and, 
although apparently free of substantial effects due to non-LTE 
or atmospheric structure (Kiselman 1993), require high S/N
($\ge$ 300) and are only detectable at relatively high metallicities.

The molecular OH lines remain an option and the near-UV electronic
OH lines were studied initially by Bessell \& Norris (1987) and
Bessell, Sutherland, \& Ruan (1991), with more recent, and quite extensive
surveys by Israelian, Garcia-Lopez \& Rebolo (1998) and
Boesgaard et al. (1999).  The region near $\lambda$3100 - 3200\AA\
is not only difficult to observe from the ground, but is also quite 
line-blanketed in stars with near-solar metallicities, and there is a
recent discussion of missing continous opacities (in the Sun) by
Balachandran and Bell (1998). Presumably, though, the importance of metal
opacities lessens as [Fe/H] decreases.   There also remains a suspicion
that the near-UV OH is sensitive to temperature structure across a stellar
disk, as well as non-LTE effects (Lambert 2001).  The use of 3D model
atmospheres may be required in order to investigate these possibilities
(Apslund 2001).  The IR vibration--rotation 
transitions of OH lie in clean regions of the spectrum and should
form in LTE (Hinkle \& Lambert 1975).   Use of the IR OH lines is still in 
an early stage (Balachandran \& Carney 1996), but with more access to
high-resolution IR spectroscopy, the number of oxygen abundance
determinations based on these lines will increase.

In order to define better the trend of B versus O at the metal-rich
end ([Fe/H] $\ge$ -0.75), oxygen abundances have been derived
for a sample of field stars in which boron abundances were
measured by Cunha et al. (2000).  These stars have 
now been observed with high-resolution and high signal-to-noise spectra
sufficient to measure either [O I] $\lambda$6300\AA\,
or the weak O I lines at 6157\AA.  Based on previous work, such as
Kiselman (1993), or the ideas summarized by Lambert (2001), these two
sets of oxygen lines should provide abundances relatively free from
non-LTE or model atmosphere effects. 
 
\bigskip
\section{Observations}

The spectra containg the neutral oxygen lines presented here were
taken with two different telescopes and spectrometers, both located
at the University of Texas' McDonald Observatory.  One subset of program
stars was observed with the Sandiford cross-dispersed echelle spectrometer, 
attached to the cassegrain focus of the 2.1m Struve reflector.
This spectrometer provides a two-pixel resolving power of 
$\lambda$/$\Delta$$\lambda$=R=60,000
on a 400 x 1200 pixel CCD detector and spectra
cover the wavelength interval 6050--7900\AA.  Nine target stars
were observed on two runs, in February and April 2000.
In addition, four other stars were observed
in September 2000 with the H. J. Smith 2.7m reflector
and cross-dispersed coude echelle spectrometer; these spectra were obtained
with a 2048 x 2048 pixel CCD at a resolution (5-pixel) of R=110,000.  
These higher-resolution data do not have complete spectral coverage and,
of the oxygen lines, only the [O I] 6300\AA\ line was visible.
Because weak oxygen lines were the primary goal of the spectroscopy, 
relatively high signal-to-noise ratios were obtained with both
instruments: S/N$\sim$ 300-400.

Data reduction utilized the IRAF software from NOAO.  Bias CCD frames
were subtracted from the raw program CCD images (stars, internal
quartz flat-fields, and Th--Ar hollow cathode lamps).  The two-dimensional
locations of the spectral orders were defined, interorder light was
then identified in each frame and polynomial fits to this light were
made in both the dispersion and cross-dispersion directions, with
the resultant interorder light subtracted from the image in question.
The bias subtracted and interorder corrected frames were then divided
by the flat-field images and the defined spectral orders were then
summed and extracted to obtain a set of one-dimensional spectra.  Wavelength
calibrations were set from the Th--Ar spectra, with typical residuals
from a wavelength fit of 5--8m\AA\ for the 2.1m spectra, and 2-3m\AA\ for
the 2.7m spectra.

In all stars, the final S/N ratios were sufficient to detect the
[O I] 6300\AA\ line and the O I lines near 6157\AA.  Because of their
weakness, and the possibilities of blending with even weaker nearby lines,
spectrum synthesis was employed in the analysis of these features.  In
addition, in the R=60,000 spectra,   
the stronger near-IR O I triplet lines, near $\lambda$7774\AA,
were easily detectable and equivalent widths measured for these lines.  
Fringing occurs
in the Sandiford CCD, redward of about 7500\AA, and this fringing
on these observing runs was not removed entirely by division by either the
flat-field or the hot star.  The wavelength scale of the fringing is
larger than the spectral-line widths, and, in particular, equivalent widths 
of the near-IR triplet lines were
measured with regard to the fringing (with amplitude of $\sim$10\%).
Comparison to published equivalent widths of these O I lines 
reveals good agreement (within 1-8m\AA), indicating that reasonably
accurate equivalent widths could be measured from these spectra for these
lines.

\bigskip      
\section{Analysis}
 
The stellar parameters (T$_{\rm eff}$, log g, and [Fe/H]) for the program
stars were the same as those adopted
in the study of boron by Cunha et al. (2000).
These parameters, in turn, were taken largely from the work of 
Boesgaard et al. (1998), with some minor modifications in the values of the
microturbulent velocities made by Cunha et al. (2000), who found somewhat
smaller values.  The effective temperatures used here
were obtained from an average of eight photometric scales
and from the infrared flux method. The log g's for the sample stars
were derived from the recipe in Edvardsson et al. (1993), which come
from a comparison of the c$_{\rm 1}$ indices
with those from the stellar models. 
Stellar parameters are listed in the first 5 columns of
Table 1.

Although it is not expected that the stellar parameters for near-solar
metallicity F and G dwarfs and subgiants are seriously in error, 
consistency checks were made on the adopted effective temperatures, gravities 
and metallicities.
In the first check, it was found that 9 of our program stars
were in common with the large abundance study of disk F and G stars
from Edvardsson et al. (1993). They derived effective temperatures from
theoretical Str\"omgren colors (primarily b--y) computed from opacity-sampled
MARCS model atmospheres and obtained metallicities from a sample
of Fe I and Fe II lines. 
A comparison of our adopted values with their T$_{\rm eff}$'s
and [Fe/H]'s revealed the following mean differences,
in the sense of (this study -- Edvardsson et al.):
$\Delta$T$_{\rm eff}$= -85$\pm$34K and $\Delta$[Fe/H]= 0.00$\pm$0.06 dex. 
Our adopted surface gravities could also be compared with the values     
in Edvardsson et al. (1993), although it was their recipe that was used in 
order to derive our log g's.
The comparison is, as expected, very
good, with $\Delta$log g= 0.00$\pm$0.04, and any differences result 
from the sensitivity of
the log g's to the different T$_{\rm eff}$'s and metallicities. 

As a further check,
our adopted surface gravities were compared with those obtained ultimately from 
trigonometric
parallaxes measured with the Hipparcos satellite. Allende Prieto \& Lambert 
(1999) derived fundamental parameters for roughly 17,000 nearby stars from 
comparison
with evolutionary calculations. A direct comparison of the log g's for
all stars studied here with the trigonometric log g's listed in their Table 1 
indicated very good agreement:
a mean difference (This Study - Trigonometric) of -0.04$\pm$0.10 dex. 
  
Another possibility was to compare the adopted stellar parameters 
with parameters determined spectroscopically.
A sample of Fe I and Fe II lines, with accurate
laboratory gf-values, were measured in all program stars from the
spectra obtained here.   A discussion of selected sample of iron lines and
their gf-values is found in Smith et al. (2000); using this set of lines
in a solar model yields an iron abundance of log $\epsilon$(Fe)=
7.50, in agreement with the meteoritic value, and a solar
microturbulent velocity of 1.0 km s$^{-1}$.  For our R=60,000 spectra,
this linelist represented 22 Fe I lines and 4 Fe II lines, while for
the R=110,000 spectra (with incomplete wavelength coverage), 5 Fe I
and 2 Fe II lines were measured.  A simultaneous fit, demanding the
same Fe abundances from low- and high-excitation ($\chi$
$\sim$2-5 eV) and weak and strong ($\sim$5-120 m\AA) Fe I lines, as
well as the Fe II lines, yields spectroscopic values of T$_{\rm eff}$,
log g, $\xi$, and Fe abundance.  In all cases for the program stars, it
is found that the Fe I and Fe II lines yield stellar parameters which
are extremely close to the adopted ones.  The mean differences (and
standard deviations) of the spectroscopic parameters minus those from
Cunha et al. are $\Delta$T$_{\rm eff}$=-25$\pm$30K, $\Delta$log g=
-0.05$\pm$0.07, and $\Delta$$\xi$= +0.2$\pm$0.2 km s$^{-1}$.  Use of
the spectoscopic parameters would have no significant effect on B and
O (or Fe), thus, in order to retain consistency with the boron results
from Cunha et al. (2000), we adopt the stellar parameters used in 
that study. 

The LTE synthesis code MOOG and
Kurucz model atmospheres were
used in order to perform spectrum synthesis for the 6157\AA\ and
6300\AA\ regions, with the linelists constructed using the Kurucz \& Bell
(1995) compilation of atomic lines, and lines other than those from
oxygen in these
spectral intervals having their respective gf-values adjusted in order to
fit the solar flux spectrum (Kurucz et al. 1984).  For the rather strong
O I near-IR lines, equivalent widths were measured and oxygen abundances 
derived from these.
Imperfect cancellation of telluric
O$_{\rm 2}$ rendered the [O I] line questionable in 3 of the 11 stars
observed at R=60,000 (and this line in these stars was not analyzed),
while in the 4 stars observed at R=110,000, the spectral coverage included
only the [O I] line.  Oscillator strengths for the permitted O I lines
at 7774 \AA\ were taken from Bell \& Hibbert (1989) and Butler \&
Zeippen (1991), while gf- values for the O I
lines at 6157 \AA\ were from Wiese et al. (1966).
The gf-value suggested by Lambert (1978) was used for the [O I]
6300 \AA\ line.
In Figure 1 are shown synthetic spectral fits to the 
weak [O I] and O I lines in HD184499: this star was chosen as an
illustration due to its 
relatively low metallicity, with the [O I] and O I lines 
easily detected and well-defined.  

The derived LTE Oxygen abundances, as well as stellar parameters and
boron abundances, are summarized in Table 1.  Abundances
from all three sets of oxygen lines, the [O I] 6300\AA\ line, the
2 O I lines near 6157\AA, as well as the near-IR O I lines near
7774\AA, are listed in columns 7--9 of Table 1.  The first 5 columns of
this table contain the stellar names and parameters, while column 6
lists the non-LTE boron abundances. 
If both the [O I] 6300\AA\ and O I
6157\AA\ sets of lines were used, the final O abundance (listed in the
last column of Table 1) is the average of both sets, otherwise, either 
6300\AA\ or 6157\AA\ is used in the final abundance. A comparison 
of abundances derived from the oxygen lines are shown in Figure 2, where
the O abundances from O I 6157\AA\ are plotted versus
those from [O I] 6300\AA: the solid line shows perfect agreement,
while the dashed lines are $\pm$0.1 dex.  These two weak-line indicators
are in good agreement, with $\sim$0.1 dex differences not at all unexpected
for such weak lines.   The mean difference in abundance from these two
sets of lines is ([O I] -- O I) +0.04$\pm$0.08 dex: this is a reasonable 
estimate of the uncertainties in the O abundances presented for this sample
of stars.  In addition, the [O I] line has an excitation
potential of $\chi$= 0.00 eV, while the 6157\AA\ O I lines have
$\chi$= 10.74 eV.  Such good agreement between lines of quite different
excitation potentials suggests that the T$_{\rm eff}$-scale used here
is adequate.  We note that Edvardsson et al. (1993) found trends between
abundances derived from the [O I] 6300\AA\ and O I 6157\AA\ lines, such that
[O/Fe]$_{\rm 6300}$= -0.025 + 0.4657[O/Fe]$_{\rm 6157}$.  The sample here is
too small, and spans too limited a range in [O/Fe], to test significantly
the Edvardsson et al. relation.  Our results, however, do fit easily within
an Edvardsson-like relation with a slope of 0.60 and an offset from them of
0.09 dex.  Again, over the limited range of [O/Fe] spanned by the stars in
this sample, the differences between [O I] and O I are less than 0.1 dex. 

The good agreement found between [O I] and O I 6157\AA\ is not found when
these weak-line O-abundance results are compared to abundances derived
from the triplet O I lines near 7774\AA. (See Table 1 as well as
Figure 1 in Cunha, Smith \& King 2001).
Attempts to find some order to the sense of the differences between
O I 7774\AA\ and the other two sets of lines do not succeed completely.
Although some trend is found with temperature, in the sense that the 7774 \AA\
lines tend to give systematically larger O abundances as T$_{\rm eff}$
increases, there is still considerable scatter at any given T$_{\rm eff}$.
King \& Boesgaard (1995) found that, in near solar-metallicity ([Fe/H]
$\ge$-0.5) F and G stars, the [O I] and O I 7774\AA\ lines show reasonably
good abundance agreements for T$_{\rm eff}$ $\le$ 6200K, but the O I lines
then yield systematically larger O abundances for the hotter stars.  This
is similar to what is found here, although we have fewer stars (note in
Table 1 that the best agreement between [O I] and O I 7774\AA\ is found
for the two coolest stars HD150680 and the Sun). 

\bigskip
\section{Results and Discussion}

It has been argued here that the O abundances derived from the
combination of [O I] 6300 \AA\ and O I 6157 \AA\ lines in this sample
of 13 field F and G stars do not suffer from significant non-LTE
or model atmosphere effects.  These abundances can be combined with
the boron abundances to define a B--O
relation for the disk.  The top panel in Figure 3 shows such a relation,
and is a plot of log $\epsilon$(B)$_{\rm NLTE}$
versus log $\epsilon$(O); the boron
abundances have uncertainties characterized by $\sim$$\pm$0.2 dex,
while we estimate here (based on the comparison of [O I] to O I 6157 \AA\
abundances in the same stars) that the O abundances have typical
uncertainties of $\pm$0.1 dex and these error estimates are indicated
by the errorbars in Figure 3.  Two stars, HD61421 and HD150680, were  
excluded from Figure 3 and from the B-O fit because these two stars exhibit
extremely large Li depletions, as well as measurable Be depletion, and thus
may also have undergone some B astration (Boesgaard et al. 1998; Cunha
et al. 2000).  The remaining 11 stars in the top
panel of Figure 3 show no evidence of B depletion.
There is a well-defined trend of
B with O in Figure 3 and, in the log-log plane, is fit by a
straight line over this (admittedly limited) range of abundance in
oxygen.   The slope of this line, derived from a linear least-squares
fit which includes the data point error estimates, is
m$_{\rm BO}$= 1.39$\pm$0.08 (the intercept of this line is -9.62$\pm$1.38).
Viewed another way, the number abundances of boron and oxygen vary
as {\bf N(B) $\alpha$ N(O)$^{1.4}$} in field disk stars.
Models which include either only direct or reverse production of B by
energetic particles predict secondary or primary behavior of B with
O: in log-log space this corresponds to slopes of either 2.0 or 1.0,
respectively.  At the metal-rich end of the metallicity distribution of
stars in the Galaxy, for this sample of field stars, 
these results rule out either pure primary or pure secondary production
of B with O, presumably indicating a combination of processes for
B production in the disk. 

It is of interest to compare the slope of log $\epsilon$(B) versus
log $\epsilon$(O) derived here in disk stars, with other studies of
boron in halo stars.  Both Duncan et al. (1997) and Garcia-Lopez et al.
(1998) have measured boron in 9 halo stars spanning the range of
[Fe/H] from -3.0 to -0.4 (for three other halo stars observed for boron,
Garcia-Lopez et al. argue only for upper limits to the B I detection,
and we discard these stars in the following discussion).  
Oxygen abundances are also available for this sample of halo stars
from the literature and both Duncan et al. and
Garcia-Lopez et al. include some comments on these literature O
abundances in their analyses of the behavior of B.  Both papers
point to the fact that oxygen abundances in halo stars, which rely
almost exclusively on the near-IR O I or near-UV OH, remain uncertain, with
large scatter found from various literature values for the same star.
In the end, these studies resort to assuming an [O/Fe] versus [Fe/H] relation
in order to estimate how B might vary with O in the halo.  It turns
out that the evolution of B with O over the entire range of halo to disk
hinges critically on the behavior of [O/Fe] with [Fe/H].  

As a first step
in investigating the possible halo behavior of boron, we used
the non-LTE boron results from both Duncan et al. and Garcia-Lopez 
et al. (for 9 stars) and simply averaged their respective values: both
studies find excellent agreement betwen their boron abundances, with
only small ($\sim$0.1 dex) differences.  Here, iron abundances will be taken
from literature values (for which there is little disagreement),
and three assumed [O/Fe] relations will be used to ''derive'' the
respective runs of boron versus oxygen.  This is the same approach as that
used by Duncan et al. (1997) and Garcia-Lopez et al. (1998) and is used
because of the uncertainty still under discussion about how the
abundance of O varies with metallicity in the metal-poor stars.  Using
this approach allows one to see quantitatively how the different possible
behaviors of oxygen in the metal-poor stars will impact the transitional
behavior of B versus O from halo to disk.  In one case, we use 
[O/Fe]= -0.4[Fe/H] over the entire range of [Fe/H] values
(as found by most of the near-IR O I studies, e.g. Cavallo, Pilachowski,
\& Rebolo 1997, or near-UV OH, e.g. Israelian et al. 1998 or Boesgaard
et al. 1999), while in the second case we assume that [O/Fe]= -0.4[Fe/H]
only down to [Fe/H]= -1.0, below which [O/Fe] stayed constant (as is
found typically by studies that use [O I], e.g. Barbuy 1988; Fulbright \&
Kraft 1999; Apslund 2001).   Use of a simple ``breakpoint'' in the [O/Fe]
relation at [Fe/H]= -1.0 is assumed, although the reality of such a breakpoint
has been investigated by King (1994).  He suggests that a simple polynomial
provides as good a fit to the data as two lines, although the differences
between the two ways of parameterizing [O/Fe] versus [Fe/H] will not
affect results discussed here.  It should also be noted that most of the
stars used to define [O/Fe] using the [O I] line are giants, and there
remains a possibility that some of these giants have undergone very
deep mixing which might mix material that is depleted in O, via the CNO
cycles, onto the surface.  Such O depletions, however, should also be
accompanied by large N enhancements, which have not been observed in
samples of field red giants (e.g. Langer, Suntzeff, \& Kraft 1992). 
We also consider the possibility that [O/Fe]
follows a gentle, but steady, increase with decreasing [Fe/H], as advocated
by King (2000), where [O/Fe]= -0.184[Fe/H] + 0.019. 
A comparison of the resulting B--O
relations are shown in the bottom panel of Figure 3 (along with the
disk relation).
It is found that for the case of a
continually increasing [O/Fe] ratio, the halo slope of log $\epsilon$(B)
versus log $\epsilon$(O) is m$_{\rm BO}$= 1.44.  This slope is
indistinguishable from the slope we derive for the disk stars, pointing
to no evolution in the yields of B/O as the oxygen abundance increases
by 2 orders of magnitude.  If, on the other hand, the constant value
of [O/Fe]= +0.4 is used below [Fe/H] of -1.0, the slope for the halo 
B abundances is then m$_{\rm BO}$= 0.92: significantly different than
what is found for the disk stars (where we argue that the oxygen 
abundances are secure).  In addition, for the steady, but gentle-sloped
increase from King (2000) we derive a similar slope of m$_{\rm BO}$= 1.05.
For these latter two cases, there would be a clear slope change 
in log $\epsilon$(B)
versus log $\epsilon$(O) near [Fe/H]$\sim$ -1.0, which would indicate
a change in yields of B/O, going from near primary in the halo, to
intermediate primary-secondary (hybrid) in the disk.  
Clearly, more work on the oxygen abundances for the halo stars for which
boron abundances have been derived is desirable.

In the above exercise, oxygen abundances have been estimated directly
from the Fe abundances using three different, assumed O--Fe relations.  The
results obtained above could be incorrect if the literature Fe abundances were
in error.  Recently, Idiart \& Thevenin (2000) have presented calculations
which indicate that Fe I may be measurably out of LTE in solar-temperature
stars at low metallicities (with the effect increasing as [Fe/H]
decreases).  We have, thus, also used the published corrections from Idiart 
\& Thevenin for Fe to rederive O abundances using their non-LTE Fe abundances
for the halo program stars.  These new oxygen estimates were then compared to
boron with the following slopes obtained: for a steadily increasing
[O/Fe] ratio of 0.4[Fe/H], a halo boron-oxygen slope is found to be
m$_{\rm BO}$= 1.57, and for the constant [O/Fe] value of +0.4 for the
halo, a slope of 1.10 is obtained, while for King's (2000) relation
the slope is 1.19.  The conclusions from the previous
paragraph remain: in one case the slope from disk to halo remains
approximately constant, while in the other two cases, there is a transition
from an intermediate slope in the disk, to primary behavior of B--O
in the halo. 

Beryllium is an interesting element to compare to boron, and
Boesgaard et al. (1999) have presented an extensive set of results
for Be versus O for [O/H] ranging from -2.0 to 0.0.  They also find
that, in a log-log plot, Be is linearly related to O with a slope
of m$_{\rm BeO}$= 1.45$\pm$0.04.  The oxygen abundances from
Boesgaard et al. are obtained from the near-UV OH, and the discussion
from the previous two paragraphs applies to this slope, as well.  
At the metal-rich
end of the distribution ([Fe/H] $\ge$ -1.0), however, there may be little to
no correction to the O abundances from the near-UV OH (Apslund 2001), thus
the results from Boesgaard et al. can be compared directly to the
results here for B versus O.  The slopes between B and Be, within the
uncertainties, are identical. 
Transformed to the log $\epsilon$
notation the Boesgaard et al. relation for Be--O is
log $\epsilon$(Be)= 1.45[log $\epsilon$(O)] - 11.57, which can be compared
to our result for boron of log $\epsilon$(B)= 1.39[log $\epsilon$(O)] - 9.62.
These two relations of B and Be versus O predict ratios of B/Be of
28 at log $\epsilon$(O)= 8.90 and 26 at log $\epsilon$(O)= 8.40: close to
the solar value of B/Be= 23 (meteoritic) and 20 (photospheric),

\bigskip
\section{Conclusions}

Oxygen abundances have been derived from weak lines of [O I] 6300\AA\
and O I 6157\AA, which are argued to be reliable indicators for
deriving O abundances, in a sample of disk F and G stars.  
The sample of disk stars for which oxygen was
measured are the same stars in which Cunha et al. (2000) have determined
boron abundances and a relation between log $\epsilon$(B)$_{\rm Non-LTE}$
and log $\epsilon$(O) has been established for these disk F and G stars.
A well-defined slope, in
log-log coordinates, is found with m$_{\rm BO}$= 1.39$\pm$0.08, which
indicates a relation intermediate between primary and secondary 
production of B with repsect to O.  This slope is identical to the
analogous slope found for Be-O by Boesgaard et al. (1999) for stars
covering the same metallicity range (they find m$_{\rm BeO}$ in a log--log
relation to be 1.45). 

A comparison of boron abundances derived here to published halo
B-abundances (from Duncan et al. 1997 and Garcia-Lopez et al. 1998)
shows good agreement in the overlap O abundance region, but the relation
of B to O in the halo still depends critically on the interpretation of
oxygen abundances derived from the near-IR O I lines or the near-UV OH:
a subject still not completely resolved.  It was shown that if the 
oxygen abundances as found from the near-UV OH lines remain, then the
intermediate slope of B versus O continues unchanged (within the
uncertainties) from disk to halo.  On the other hand, if O abundances,
as derived from the near-UV OH, are lowered, by $\sim$-0.25 dex at
[Fe/H]=-2.0 and $\sim$-0.50 dex at [Fe/H]=-3.0 (Apslund 2001), the 
behavior of B with O
in the halo becomes nearly primary, indicating a change in the B--O
relation from halo to disk (with the slope increasing from $\sim$ 1.0 to
1.4).

We thank the staff of McDonald Observatory for maintaining excellent 
spectrometers.  This research is supported in part 
by NASA through contract NAG5-1616, and the grant GO-06520.01.95A from the
Space Telescope Science Institute, which is operated by the Association of
Universities for Research in Astronomy, Inc., under NASA contract NAS5-26555.  
We also acknowledge support from the National Science Foundation through 
grant AST99-87374.

\clearpage

\figcaption[Cunha.fig1.ps]{Spectra of one of the more metal-poor 
sample stars: HD184499. The filled circles are the
observed points and the solid curves are the synthetic spectra. Two
spectral regions are illustrated: the upper panel
shows the forbidden line, [O I] $\lambda$6300\AA, while the bottom panel
shows the excited, permitted O I lines near $\lambda$6157\AA. These weak
oxygen lines provide O-abundances that are relatively free from 
non-LTE effects and are the primary oxygen abundance indicators for
all studied stars.
\label{fig1}}

\figcaption[Cunha.fig2.ps]{A comparison of oxygen abundances derived from
the forbidden line at $\lambda$6300\AA\ and
the weak excited O I lines at $\lambda$6157\AA\ and $\lambda$6158\AA.
There are 6 program stars in which both sets of lines
were measured, as well as results for the Sun from the Solar Flux Atlas
of Kurucz et al. (1984). The solid line depicts a perfect agreement,
while the dashed lines illustrate $\pm$0.1 dex from perfect agreement.
These results demonstrate that, together, the [O I] $\lambda$6300\AA\ line
and the O I $\lambda$6157\AA\ and $\lambda$6158\AA\ lines yield 
completely consistent results, within $\sim$0.1 dex, over the range of
T$_{\rm eff}$'s, gravities, and oxygen abundances spanned by the program
stars.
\label{fig2}}

\figcaption[Cunha.fig3.ps]{The top panel shows the 'metal-rich' disk relation
between boron and oxygen for stars studied here: all of these program stars 
have undepleted B. The oxygen abundances are based 
exclusively on the weak [O I] $\lambda$6300\AA\ and $\lambda$6157\AA\
lines. The non-LTE boron abundances are from Cunha et al. (2000).
In the log-log plot of abundances, there is a well-defined 'linear'
relationship (correlation coefficient of 0.90) with a slope of
1.39$\pm$0.08, or N(B) proportional to N(O)$^{1.4}$. We expect
this relationship, defined at the metal-rich end, to be free
from significant systematic effects.  The bottom panel compares the
relation defined for the disk stars to boron abundances obtained for
halo stars from Duncan et al. (1997) and Garcia-Lopez et al. (1998).  Because
oxygen abundances from the near-UV OH lines and near-IR O I triplet lines
(which are the lines used in the halo stars)
remain uncertain, oxygen abundances are estimated from Fe abundances using
three different assumptions about the behavior of O with Fe in the halo. 
The m$_{\rm BO}$= 0.93 line is derived from an O/Fe relation in which
[O/Fe] is constant (at +0.4) below [Fe/H] of -1.0, as suggested in the
recent preprint by Apslund (2001--with the sample data
points being the open squares).  The m$_{\rm BO}$= 1.05
line is from a King (2000) relation of [O/Fe]= -0.184[Fe/H] + 0.019 (with
the data indicated by the filled triangles),
while the m$_{\rm BO}$=
1.44 line results from [O/Fe] increasing steadily as -0.4[Fe/H], as indicated 
by the OH results from Israelian et al. (1998) and Boesgaard et al. (1999--
with the data shown by the crosses).
\label{fig3}}

\end{document}